\documentclass[10pt,twoside]{article}
\usepackage{graphicx}
\usepackage{psfig}
\newcommand{\be}{\begin{equation}}
\newcommand{\ee}{\end{equation}}

\makeindex

\begin{document}

\title{Complexes of Stars and  Complexes of Star Clusters}


\author{Yu.N.\,Efremov\\
\it Sternberg Astronomical Institute, Moscow, Russia\\
\it e-mail: efremov@sai.msu.ru\\}

\date{~}
\maketitle

\thispagestyle{empty}

\begin{abstract}
\

     Most star complexes are in fact  complexes of stars, clusters
and  gas clouds; term "star complexes" was introduced
as general one disregarding the preferential content of a complex.
Generally the high rate of star formation in a complex is accompanied by
the high number of bound clusters, including massive ones,
what was explained by the high
gas pressure in such regions. However, there are also complexes, where
clusters seems to be more numerous in relation to stars than in a common
complex. The high rate of clusters - but not isolated stars -
formation seems to be typical
for many isolated bursts of star formation, but  deficit of stars
might be still explained by the observational selection. The latter cannot,
however, explain the complexes or the dwarf galaxies, where the high
formation rate of only stars is observed.  The possibility
of the very fast dissolution of parental clusters just in such regions
should itself be explained. Some difference in the physical conditions
(turbulence parameters  first of all) within the initial gas supercloud
might be a reason for  the high or low stars/clusters  number ratio in
a complex.

\end{abstract}

\section{Introduction}

     The common opinion is that  formation of star clusters is
the only way of star  formation. Since 1950s we
know that the stars forms by groups, in clusters and
associations.  The presently isolated stars might be formed  in
groups, which have been dissolved till the present time.

     Most star complexes are in fact  complexes of stars, clusters
and  gas clouds; term "star complexes" was initially suggested
for all of them, disregarding what is their preferential or more evident
population [1, 2]. However, later on we have presented evidences
for existence of complexes, where   clusters seems to be more numerous
in relation to stars than in a common complex [3].  Some difference
in the physical conditions within the initial gas supercloud
might be a reason for such high C/S number.

    We present here more examples of locations where the rate of
cluster formation seems do not correlate well with the rate
of the isolated star formation. These cases are to be studied in
more details. The high C/S ratio might be explained by the observational
selection, but it is more difficult for the low C/S ratio.
The plausible explanation of low C/S ratios by the very fast dissolution
of parental clusters just in such regions should itself be explained.

\section{The young massive clusters}

      The related issue is what are conditions which lead
to formation of the young massive clusters (YMCs).
The most massive of the YMCs  are bound, being the
the progenitors  of the classic (old) globular clusters for the future
times. In other wording, observed now bound YMCs looks like it had have
to be the case for the classic present day globulars some 10 Gyr ago.

     The results of the  systematic search for the YMCs, performed by
Larsen and Richtler [4]  in 21 spiral galaxies lead these authors
to conclusion that the number of such clusters (normalized to the
luminosity of the parent galaxy) correlates with
the star-formation rate (SFR) per unit of area of the corresponding
galaxy.  These authors concluded that the formation of numerous
YMCs in interacting and starburst galaxies can be explained
by the same mechanisms as in the case of normal galaxies, but operating
under extreme conditions. Whitmore [5]  corroborated this conclusion
and found that the relation between the SFR and the occurence
of YMCs found by Larsen and Richtler [4]  can be extended
to these galaxies.  These results seems to be the strong  confirmation
of the conclusion that the general high SFR is  going along with
the high occurence of star clusters.

    There are well based theoretical reasons for  the formation
of massive bound clusters in the regions of the high SFR, because
there the high pressure  conditions should exist ([6], [7], [8].
The high pressure (density) surroundings may prevent the dissolution
of even very massive newborn clusters, in spite of their O-stars
and SNe pressure to the intracluster gas. This mechanism is evidently
in action within the bursts of star formation in the interacting
galaxies, and in the regions of the high SFR generally, like it
was shown in [4] and [5].

     There are however the important exclusions. A few of the BCD galaxies
host  very luminous YMC (known as star superclusters, SSC),
which are far above the relation between  the SFR and the luminosity of
the brightest in a galaxy cluster [9]. Something else was probably in action.

     The lower abundance which is often observed  in the dwarf
galaxies might act to the same  direction; however, the
difference in the parameters of the turbulence
in the insterstellar gas may be the most important factor  to determine
the mode  of the star formation. The  latter is presumably a fast process,
determined mainly by  the interplay  of gravitation and turbulence
(Elmegreen  [10]) and the fast free
collapse of molecular cloud is prevented not by magnetic field
but by turbulence  (see, however, Muchovias [11]).
Its parameters might determine whether
the bound or fast dissolving clusters form predominantly.

     Klessen et al. [12, 13] suggested that the
decaying or long-wave turbulence results in fast
formation of bound clusters, whereas the isolated field stars form in the
case of shorter wavelengths.  Inefficient and isolated star formation
occur in such regions, but the velocity structure in molecular clouds
is dominated  by large scale turbulent modes.  If the self-gravity
overwhelms turbulence, due to compression by a large-scale shock or
to fast decay of turbulence, the bound clusters are formed. A small scale
turbulence requires un unrealistically large
number of driving sources, what explains why most stars in
the Galaxy formed in open clusters, as found in  [14].

    If so, the properties  of turbulence
might be different within the whole star/cluster complex (or even in
a draft galaxy) and this may be able to explain the different S/C
ratios in different regions.

\section{Complexes of stars and complexes of clusters}

      There are  indeed the rare complexes of only clusters  and complexes
of only isolated stars;  the  most striking  samples  are known
in the LMC. Inside the  group of a dozen coeval clusters around
NGC 2164 (four of these are the YMCs) in the LMC only three
Cepheids are known (and a few  more inside clusters), although
this complex has an optimum age for Cepheid stage of massive
star evolution  (fig. 1).

     While searching for the local agent responsible for formation
of either star clusters or isolated stars, it is important to deal
with the objects of the same age. The Cepheid stars  are easy
detectable and their ages are known from the period - age relation
well established both from observations [15]  and the  theory [16].
An objective comparison of the distributions of Cepheids and clusters of the
same age as Cepheids in the LMC  yielded, apart from NGC 2264 group,
three more groups of clusters, and only one  of these four groups
of clusters was found to coincide with a Cepheid concentration
(Battinelli and Efremov [17]). It follows from the results of  the
OGLE program (search for the variable stars) that this group, located
immediately at the East tip of the LMC bar
around the young massive massive clusters NGC 2058 and NGC 2065,
contains about twenty smaller clusters and about 150 Cepheids of
which 20 are cluster members. This complex might be considered
to be normal in cluster/star ratio.

    Immediately southeast of this NGC 2058/2065 complex there is a dense
group of about the same size  (200 x 300 pc), which hosts ~180
Cepheids and no conspicuous clusters. The central part of this
group is devoided of any clusters completely, whereas the density
of Cepheids there is about 900 per kpc$^{-2}$, i.e., by two orders
of magnitude higher than in the solar neighborhood. The periods,
and consequently, the ages of most of these Cepheids are confined
within a narrow interval (3--5 days). It follows from the age
spread and stellar density that this complex of Cepheids is a
relic of a star-formation burst - some 50-100 Myr ago there was
indeed the very high local SFR. However, this did not lead to
formation of star clusters  - or, at least, only small and fast
dissolving clusters were formed there (fig. 2).

    The  compact groups of clusters are seen also in the ACS HST  image
of the M51 galaxy. Some of them are within the spurs, which might
be the essential clue to seemingly unnormal C/S ratio  within
these groups (fig. 3). This ratio is to be studied yet with data
on the fainter stars.

\section{Discussion}

     The superassociatons or the localized bursts of star formation
are nothing but the star/cluster complexes, involved  totally
in the intensive star formation process; most of them are the supergiant
HII regions. However, it is not the case for NGC 205 = OB 78
superassociation in M31.  This ~1 kpc in size region is  filled
with OB-stars; a few small HII regions are near, but outside, of it (fig. 4).
This was explained long ago by the lack of the gas inside OB78 -
the gas was therefore expulced??? from the superassociation  by the
pressure from O-stars/SNe earlier than  it was ionized.
This plausibly  was due to the high density
of these stars; this suggestion might be checked by the observational
data on stars and gas in this  region - and also in M51, where
the very various interrelations between HII gas and star/cluster complexes
are seen and are to be studied.  Anyway, our present goal is to
stress the absence of star clusters inside OB 78. This is  an evident
contradiction to the high pressure explanation  of the formation
of the bound massive clusters. Moreover, this case is  dissimilar
to many other localized  bursts of star formation, where just the YMCs,
if not SSCs, dominate, like it is the known case in the Antennae galaxies,
and, to lesser extension, inside the M101 supergiant HII regions.

     Somewhat similar case is IC 10, the dwarf galaxy of the Local
group.  The  current  rate of star formation there is the highest of the all
Local Group galaxies, but no YMCs are known there  (Grebel, [18]).
As S.Larsen has commented, "may be there is someting peculiar going on
there" ([18], p. 427).

     The very low  cluster formation rate is known in the
irregular Local Group galaxy IC 1613. Being  normalized to the
same star-formation rate, it is 600 times lower than in the LMC, which is
a galaxy of the same type  [19].
The quite low number of star clusters in IC 1613 has been known since
W.Baade's investigations in Fifties. This led Hodge (Galaxies, 1986)
to conclusion that the unrecognised agent should exist, determining
whether the formation of the populous clusters is possible in a galaxy
or not. Anyway, 13 OB-associations and even a superassociation,
noted by W.Baade, are  known in IC 1613.  Therefore, it looks like
this agent is not always the dierect consequence of the general SFR.

    There may exist not only local, but also temporal difference
in the clusters and stars rate of formation. At least in the LMC
it is the case. The break in the formation of (at least massive)
clusters, which  lasted in the LMC for 4--14 Gyr, was not matched
by a decrease in the star-formation rate (van den Bergh [20]).
The conclusion by van den Bergh was:
"The dramatic contrast between the history of the cluster formation
and that of field stars suggests that star clusters cannot be used as
proxies for star formation".

     The steep IMF found by Massey [21] for the seemingly isolated
stars of the LMC field might be considered the indication of different mode
of formation for isolated stars, but recently  Elmegreen and Scalo [22]
concluded that this steeper IMF resulted from unappropriate assumption
of the constant SFR, which is hardly is justified for the field stars.

     All in all, these data demonstrate  the important
and often unrecognised
problem does exist  and should be carefully investigated.
The best approach to this issue  seems to be studying the
populations of the same age within isolated star/cluster complexes in
resolvable galaxies.
(The data on the MW galaxy suffered from many selection effects and
uncertainty of distances; anyway, they suggest also  the existence
of complexes of clusters,  like the compact group in Cassiopeia,
described in [3], p. 204).

     The objects within a complex have arised from the (initially) bound gas
supercloud and the ISM properties within  it must be reflected in
stars/clusters number ratio  over all the complex. This ratio
for the objects of the same
age should be determined for  the large number of the best outlined
complexes  in different galaxies and at different location within
a galaxy.  The Cepheid investigations would be valuable,
yet considering these are time-consuming and age-limited,
the data on the  bright stars could be used up to
the same magnitude limit  in a galaxy.

    The position of a complex in a galaxy   might be  connected
with the value of stars/clusters ratio inside the complex.
There are indeed some guesses that this might be the  case.
OB78 in M31 seems to locate at the cross  of two spiral arms going
to opposite directions. In the LMC, the NGC 2164 group of clusters
is in isolated position  far from the LMC center, whereas  the
NGC 2058/2065 group is near the tip of the bar and the dense group
of Cepheids  is to SW of it.   The  bright complexes
are often at the tips of  the spiral arms, whereas  in the grand
design galaxies complexes are often  located at equal spacing along the
arm,  suggesting the  formation under action of Parker instability.
The different mechanisms of formation may be reflected in the
different clusters/stars ratio and parameters of turbulence in ISM might
depend on a position inside a galaxy.

    The peculiar shape may also be connected with the certain mechanism
of formation and C/S ratio.  The arc-shaped complexes might be formed
in result of the dynamical pressure.
The arc-like shape has  been observed too at the all-galaxy scale,
for the leading edges of a number of galaxies moving through the IGM,
i.g. for DDO 265 in M81 group (fig. 5). More examples are given in  [23].
Note that the  arcs of Sextant in the LMC and that of the Western
complex in M83 consist  of 5 - 7 clusters each,  whereas the older arc of
Quadrant  consists from  both clusters and stars [24].
The star formation triggered by the high (especially dynamical?)
pressure may result mostly in the bound clusters, whereas gravitational
fragmentation of a supercloud probably lead to the normal
(most often observed) C/S ratio. This is surely the most frequent case
of a star/cluster complex formation. However, the low C/S ratio
in regions presumably  formed under the high pressure conditions,
which  demonstrate the high SFR, seems to be not only rare,
but also enigmatic cases.

    The issue of the mode of star formation might have the deep cosmological
implications. For example, the formation of the YMC (and more so of the
SSC) in the BCD galaxies, many of which are isolated, miight be
triggered by gas infalling in dark-matter haloes; this gas may experience
sloshing  and oscillation favoring compressiion and instabilities [25].

    This work has been done with the support of Russian Foundation
for the Basic Investigations, project RFFI 03-02-16288.

\section*{References}

1. Yu.N.Efremov. Sov. Astron. Lett.        1978.

2. Yu.N.Efremov. Astron. J., 110, 2757, 1995.

3. Yu.N.Efremov. Centres of Star Formation in Galaxies:
      Star Complexes and Spiral Arms. Moscow, Nauka Publ.,
      248 pp. (in Russian), 1989.

4. S.S.Larsen, T.Richtler. Astron. Astroph., 354, 836, 2000

5. B.C.Whitmore, in  "Extragalactic star clusters", ASP IAU Symp. Ser.207,
       2002, eds. D.Geisler et al., p. 367.
6. C.J.Jog, P.M.Solomon, Astrophys. J., 276, 114, 1984.

7. B.Elmegreen., Yu.N.Efremov, Astrophys. J., 480, 235, 1997.

8. B.Elmegreen., Yu.N.Efremov, R.E.Pudritz, and H.Zinnecker, in
    Protostars and Planets IV, p. 179, U. Arizona Press, 1999.

9. S.S.Larsen, in  "Extragalactic star clusters", ASP IAU Symp. Ser.207,
      2002, eds. D.Geisler et al., p. 421.

10. B.Elmegreen, Astrophys. J., 530, 277, 2000.

11. T.Ch.Mouschovias, K.Tassis, M.W.Kunz, Astro-ph/0512043, 2005.

12. R.S.Klessen, F.Heitsch, M.-M.MacLow, Astrophys. J., 535, 887, 2000.

13. R.S.Klessen, Astro-ph/0301381, 2003.

14. F.C.Adams,  Ph.C. Myers, Astrophys. J., 553, 744, 2001.

15. Yu.N.Efremov, Astronomy Rep. 47, 1000, 2003.

16. G.Bono, M.Marconi, S,Cassisi et al., Astrophys. J., 621, 966, 2005.

17. P.Battinelli, Yu.Efremov, Astron. Astroph., 346, 778, 1999.

18. E.Grebel, in  "Extragalactic star clusters", ASP IAU Symp. Ser.207,
        2002, eds. D.Geisler et al., p. 94.
19. T.K.Wider, P.W.Hodge, A.Cole, Publ. ASP, 112, 594. 2000.

20. S. van den Bergh S. in New views of the Magellanic Clouds.
     IAU Symp. 190, ASP Conf. Ser., eds. You-Hua Chu et al., p. 573, 1999.
21. P.Massey, Astroph.J., 141, 81, 2002.

22. B.G.Elmegreen, J.Scalo, Astro-ph/0509282, 2005.

23. Yu.N.Efremov, A.D.Chernin, Physics-Uspekhi, 46, 1, 2003.

24. Yu.N.Efremov, Astron. Zh., 79, 879, 2002 = Astron. Rep.,46, 791, 2002.

25. F.Combes, Astro-ph/0410410, 2004.


\section*{Fugure captions}

Fig. 1. The group of YMCs around NGC 2164 in the LMC. DSS image.

Fig. 2. a) The DSS image of the LMC field  near the bar Eastern tip,
         including the NGC 2065 and NGC 2058 clusters at the upper right
         and the region of the highest density of Cepheids at bottom left.
        b) The map of the same region, constructed with the programe
           OGLE data. The crosses are Cepheids.

Fig. 3.  a) The largest spur in M51, starting at the West of the
            spiral arm. It includes large star clusters.
         b) The detail of the fig. 3a - group of the star clusters
            located near the starting point of the spur.
         c) The star clusters inside the SW  spur, starting from
            the same M51 arm.
            It looks like these positions are favorite for the formation
            of bound clusters.  The images are details of the ACS HST image.

Fig. 4.  The bright complex (localized starburst) NGC 205 = OB78
         located at the crossing of the spiral arms S4 and  S3 in M31.

Fig. 5.  The dwarf galaxy DDO 265 in M81 group.  Image taken at the BTA.


\end{document}